\documentclass[aps,prb,twocolumn,groupedaddress,showpacs,superscriptaddress,amssymb,amsmath]{revtex4-1}
\usepackage{graphicx}
\usepackage{tabularx}
\usepackage{color}
\usepackage{amsmath}
\usepackage{comment}
\newcommand{\be}{\begin{equation}}
\newcommand{\ee}{\end{equation}}
\newcommand{\bea}{\begin{eqnarray}}
\newcommand{\eea}{\end{eqnarray}}
\usepackage{dcolumn}
\usepackage{hyperref}
\usepackage{bm}
\usepackage{epsf}
\usepackage{subfigure}
\usepackage{amsfonts}

\begin{document}
\title{Thermodynamic uncertainty relation in thermal transport}
\author{Sushant Saryal}
\affiliation{Department of Physics, Dr.~Homi Bhabha Road, Indian Institute of Science Education and Research, Pune, 411008 India}
\author{Hava Friedman}
\affiliation{Department of Chemistry and Centre for Quantum Information and Quantum Control,
University of Toronto, 80 Saint George St., Toronto, Ontario, Canada M5S 3H6}
\author{Dvira Segal}
\affiliation{Department of Chemistry and Centre for Quantum Information and Quantum Control,
University of Toronto, 80 Saint George St., Toronto, Ontario, Canada M5S 3H6}
\author{Bijay Kumar Agarwalla}
\affiliation{Department of Physics, Dr. Homi Bhabha Road, Indian Institute of Science Education and Research, Pune, 411008 India}

\date{\today}
\begin{abstract}

We use the fundamental nonequilibrium steady state fluctuation symmetry
and derive a condition on the validity of the thermodynamic uncertainty relation (TUR)
in thermal transport problems, classical or quantum alike.
We test this condition and study the breakdown of the TUR in different thermal transport junctions
of bosonic and electronic degrees of freedom.
First, we prove that the TUR is valid in harmonic oscillator junctions.
In contrast, in the nonequilibrium spin-boson model, which realizes many-body effects,
it is satisfied in the Markovian limit, but 
violations arise as we tune (reduce) the cutoff frequency of the thermal baths, 
thus observing non-Markovian dynamics.
Finally, we consider heat transport by noninteracting electrons in a tight-binding chain model. Here we show that
the TUR is feasibly violated by tuning e.g. the hybridization energy of the chain to the metal leads.
These results manifest that the validity of the TUR relies on the statistics of 
the participating carriers, their interaction,
and the nature of their couplings to the macroscopic contacts (metal electrodes, phonon baths).
\end{abstract}
\maketitle

\section{Introduction}
\label{sec-intro}
The thermodynamic uncertainty relation (TUR) 
 \cite{Barato:2015:UncRel,Gingrich:2016:TUP,Polettini:2016:TUP,Pietzonka:2016:Bound,Hyeon:2017:TUR,Horowitz:2017:TUR,Proesmans:2017:TUR,Pietzonka:2017:FiniteTUR,Garrahan:2017:TUR,Dechant:2018:TUR, Pietzonka:2017:FiniteTUR,Falasco,SamuelssonM,Garrahan:2017:TUR,Dechant:2018:TUR,Koyuk:2018:PeriodicTUR,Garrahan18,Gabri,Vu,Timpanaro,Hasegawa1,Hasegawa2,Saito,TUR-gupta, TURQ,BijayTUR,JunjieTUR,SamuelssonQP}
a recently discovered trade-off in the field of stochastic thermodynamics \cite{St-thermo,St-thermo2}, 
provides a low bound for the relative uncertainty of observables in terms of the associated 
entropy production. The precise statement of the TUR for a two-terminal out-of-equilibrium system 
in steady state reads
\bea
\frac{\langle  j^2\rangle_c}{  \langle j\rangle^2} \frac{\langle \sigma \rangle }{ k_B} \geq 2.
\label{eq:TUR0}
\eea
Here, $\langle j \rangle$  is the averaged current, $\langle j^2 \rangle_c$ is the second cumulant, and  
$\langle \sigma\rangle$ is the averaged entropy production rate, all evaluated in steady state;
$k_B$ is the Boltzmann constant. 

The TUR, Eq.~(\ref{eq:TUR0}), was put forward in Ref. \cite{Barato:2015:UncRel} 
for biomolecular processes in the linear response regime. 
A generalized version of the TUR had been proposed and tested for various systems 
under driving \cite{Pietzonka:2017:FiniteTUR,Garrahan:2017:TUR,Dechant:2018:TUR,Falasco,SamuelssonM,Vu}.
Nevertheless, generalized TURs (see e.g. Ref. \cite{Vu,Timpanaro}) that are given in terms of the total entropy production
are inconsequential in the steady state limit. 
Therefore, for systems operating under a constant voltage or thermal bias in the steady state limit
one should focus on the TUR as given by Eq. (\ref{eq:TUR0}).
The validity of this relation, so far, has been established for classical markovian dynamics.
While one expects that a related fundamental bound would exist in general cases,
the validity of Eq. (\ref{eq:TUR0}) in the quantum regime \cite{Goold}, 
as well as for systems beyond the markovian limit, is questionable.
In our previous work \cite{BijayTUR}, we explored the TUR in charge transport problems, which are quantum in nature,
and demonstrated its violations in quantum dot junctions of noninteracting electrons.
In Ref. \cite{JunjieTUR}, we further examined the validity of the TUR in thermoelectric junctions of noninteracting electrons,
and analyzed the related power-efficiency-power fluctuation trade-off relation.

What is the origin of the TUR? Can we derive it from fundamental principles?
Using the nonequilibrium steady state fluctuation symmetry (SSFS), we put together
the relative uncertainty of heat current and the associated entropy production rate.
The resulting expression recovers the structure of the TUR---albeit allowing its violation---depending on the 
sign of high order cumulants.
This analysis, based on the SSFS, allows us to interrogate the 
TUR in thermal energy transport problems along several axes:
classical-quantum dynamics, harmonic-anharmonic systems, bosonic-fermionic statistics for the participating particles, and Markovian-non Markovian dynamics.
The role of interaction on the breakdown of the TUR is examined by contrasting a fully harmonic model
to the spin-boson model. Furthermore, we study here
bosonic, fermionic, and hybrid statistics (spin-boson) systems, illustrating
that the particle statistics plays a central role on this bound.
Quantum effects are further questioned by analyzing the TUR at high 
temperature for both bosonic and electronic systems.
Finally,  the role of non-Markovian effects on the TUR is recognized
by controlling system-bath parameters such as the spectral density function of the 
attached (phononic or electronic) thermal baths.


Our setup includes a central system that is coupled to two heat baths $L,R$ that are maintained at different 
temperatures, $T_L>T_R$;
there are no other thermodynamical forces (e.g. the chemical potential is fixed).
For such a setup, the average steady-state entropy production rate is given by
$\langle \sigma \rangle= k_B\Delta \beta \langle j\rangle$, 
where $k_B\Delta \beta= \frac{1}{T_R}-\frac{1}{T_L}$ is the thermodynamic affinity,  
driving the system out-of-equilibrium in an irreversible manner. 
In this setting, the TUR (\ref{eq:TUR0}) simplifies to
\bea
\Delta \beta \frac{\langle j^2 \rangle_c}{  \langle j\rangle} \geq 2.
\label{eq:TUR1}
\eea
%
The objective of this this work is to understand the behavior of current
fluctuations in thermal transport junctions, by providing insights on 
the validity of the TUR, Eq. (\ref{eq:TUR1}).
First, based on the steady state fluctuation symmetry we show that TUR violations are 
linked to the behavior of the skewness, the third cumulant of the heat current. 
This derivation holds for both classical and quantum systems.
We exemplify this observation and examine the TUR in three central quantum thermal transport models:
(i) Chain of coupled harmonic oscillators. In this case we prove that the TUR is always satisfied.
(ii) Nonequilibrium spin-boson model. Here, we demonstrate that the TUR can 
be violated by structuring the thermal baths, which leads to non-Markovian dynamics.
(iii) Fermionic chains, where we study electronic heat transport. 
In the resonant transport regime we write down an analytic condition for TUR violations, which is
supported by numerical simulations.

The paper is organized as follows. 
In Section \ref{Sec-bos}, we employ the universal steady state fluctuation relation                 
(Gallavotti-Cohen symmetry) for heat exchange and
present a $\Delta \beta $ perturbative expansion of the TUR.
In Section \ref{Sec-har},  we study the non-interacting harmonic oscillator model.
The nonequilibrium spin-boson (NESB) model is examined in Sec.  \ref{Sec-SB}.
In Section \ref{Sec-ele} we investigate the TUR in electronic heat transport. 
Certain details are delegated to the Appendices.
We conclude and discuss future directions in Section \ref{Sec-summ}.

\section{Perturbative expansion for the TUR from fluctuation symmetry}  
\label{Sec-bos}

Invoking the steady state fluctuation symmetry for heat exchange \cite{Esposito-review,fluc1,Pal},  
we derive here a perturbative expression for the ratio $\Delta \beta \langle j^2\rangle_c/\langle j\rangle$
in orders of the affinity $\Delta \beta$.
In steady state, we can formally expand the averaged heat current and its higher order cumulants, 
$\langle  j^n \rangle_c$, $n$=1, 2, 3,... in powers of the affinity $\Delta \beta$ as
\bea
&&\langle j \rangle = G_1 \Delta \beta + \frac{1}{2!} G_2 (\Delta \beta)^2 
+ \frac{1}{3!} G_3 (\Delta \beta)^3 + \frac{1}{4!} G_4 (\Delta \beta)^4+ \cdots \nonumber \\
&&\langle j^2 \rangle_c=S_0 + S_1 (\Delta \beta) + \frac{1}{2!} S_2 (\Delta \beta)^2 + \frac{1}{3!} S_3 (\Delta \beta)^3+ \cdots \nonumber \\
&&\langle j^3 \rangle_c = R_1 \Delta \beta +  \frac{1}{2!} R_2 (\Delta \beta)^2+ \cdots \nonumber \\
&&\langle j^4 \rangle_c = T_0  +  T_1 (\Delta \beta)+  \cdots 
\hspace{-5cm} \label{eq:transport}
\eea
The four cumulants are the averaged current, its variance, skewness, and the kurtosis.
For the average current, $G_1$ is the linear transport coefficient, or the thermal conductance,
$G_2,G_3, \cdots $ are nonlinear transport coefficients. 
Heat current fluctuations include the equilibrium noise component
$S_0$ and higher order nonequilibrium terms,  $S_1$, $S_2\cdots$.
Other coefficients appearing in the skewness and the kurtosis, such as $R_1, R_2, T_0, \cdots$ 
can be similarly described. 

The steady state fluctuation symmetry relates the probability of integrated heat current,
$Q(t) = \int_{0}^{t} d\tau j(\tau)$, flowing in the forward direction (from hot to cold) $p(Q=j t)$ to that of the probability of heat flowing in the reverse direction $p(-Q=-j t)$ (from cold to hot). 
The precise statement of SSFS is
\begin{equation}
\lim_{t\to \infty} \frac{1}{t} \ln \Big[ \frac{p(Q=j\,t)}{p(Q=-j\,t)}\Big]= \Delta \beta j.
\end{equation}
Here, $\sigma= k_B\Delta \beta j$ is the stochastic entropy production rate.  
In terms of the generating function defined as the Fourier transform of the probability distribution, 
$Z(\alpha) \equiv \int dQ e^{i \alpha Q} p(Q)$, the SSFS leads to the symmetry 
$Z(\alpha) = Z(-\alpha + i \Delta \beta)$. 
Here, $\alpha$ is the counting parameter, which keeps track of the net amount of thermal 
energy flowing between a bath to the system. 
The cumulants for current are obtained by taking derivatives,  
$ \langle j^n \rangle_c = \frac{\partial^n \chi(\alpha)}{\partial (i\alpha)^n} \Big|_{\alpha=0}$, 
where $\chi({\alpha}) = \lim_{t\to \infty} \frac{1}{t} \ln Z(\alpha)$ is the  cumulant generating function (CGF) for heat transport. 

As a consequence of this SSFS, it can be shown that 
linear and higher order transport coefficients in Eq. (\ref{eq:transport}) 
are in fact related to each other \cite{fluct3},
\bea
&&S_0 = 2\,G_1, \quad S_1 = G_2, 
\nonumber \\
&& T_0 = 2\, R_1, \quad T_1 = R_2, 
\nonumber \\
&&3 S_2 - 2 G_3 = R_1, \nonumber \\
&&2 S_3 - G_4=R_2, \nonumber \\
&&\cdots
\label{eq:sym-relations}
\eea
and so on. 
We substitute the expansion for the current and its fluctuations Eq. (\ref{eq:transport})
into Eq.~(\ref{eq:TUR1}) and simplify it using the relationships (\ref{eq:sym-relations}). As a result, we
obtain the left hand side of the TUR as a perturbative series in $\Delta \beta$,
\bea
\Delta \beta \frac{\langle j^2 \rangle_c}{\langle j \rangle}&=& 2 +  \frac{(\Delta \beta)^2}{6 G_1} \, R_1
+ \frac{(\Delta \beta)^3}{12 G_1^2} \, \Big[G_1 R_2 - G_2 R_1\Big] \nonumber \\ &+& \mathcal{O}{(\Delta \beta)^4} + \cdots
\label{eq:central-eq}
\eea
This expression is one of the central result of the paper. 
Since there is no fundamental constraint on the sign of the skewness coefficient, $R_1$,
the TUR (\ref{eq:TUR1}) may be violated in nonequilibrium systems---within a certain range of the affinity 
$\Delta \beta$.

We point out the following:
(i) The expansion (\ref{eq:central-eq}) is valid for both classical and quantum systems. 
(ii) The linear order term,  $\Delta \beta$, does not contribute to the TUR. 
This is a consequence of the relation $S_1 = G_2$,
which stems from the time-reversal symmetry of the underlying Hamiltonian. 
(iii) Coefficients for orders $(\Delta \beta)^n$, $n=2,3 \cdots$  include  heat current cumulants 
that are greater than two. 
Therefore, if the probability distribution for heat exchange is Gaussian 
the TUR precisely saturates to the value $2$, since all higher order cumulants greater than two vanish. 
(iv) It is evident from this expression that for noninteracting setups (missing a diode effect) only 
even orders in $\Delta \beta$ contributes since
the current is an odd function of the thermal bias, while the second cumulant involves only
 even powers in $\Delta \beta$. 
This may not be the situation for interacting junctions.

Equation (\ref{eq:central-eq}) reveals that the TUR (\ref{eq:TUR1}) is violated if  $R_1<0$. 
Of course, one may observe that $R_1>0$ but the TUR is still violated due
to the impact of higher cumulants. 
However, in this work we examine the breakdown of the TUR within the leading order of the affinity i.e., up to 
$(\Delta \beta)^2$, and we therefore focus on the sign of $R_1$.
In what follows we study the TUR in three representative thermal transport problem: harmonic oscillator system
coupled to harmonic baths, a central spin coupled to harmonic baths, and a tight-binding electronic junction. 

A recent study obtained a current fluctuation - entropy production trade-off relation based on the geometry of 
quantum non-equilibrium steady-states \cite{Goold}. This quantum TUR (QTUR) reads
 $\langle j^2\rangle_c \langle \sigma\rangle/(k_B\langle j \rangle^2)\geq 1$, which is two times looser than
Eq. (\ref{eq:TUR0}) [or Eq. (\ref{eq:TUR1}) for the specific case of heat transport].
Of course, once a system obeys the QTUR, it satisfies as well the `standard' TUR (\ref{eq:TUR0}).
We also emphasize that while we show below violations to the bound (\ref{eq:TUR1}), 
our results do obey the QTUR. 
Nevertheless, we argue that analyzing the TUR (\ref{eq:TUR0}) 
is particularly interesting for several reasons:
First, at equilibrium the TUR becomes an equality, 
providing a clear reference point to as the role of the nonequilibrium condition on the trade-off relation.
Further, the expansion (\ref{eq:central-eq}) builds about the
equilibrium value of 2, and it provides insights on
violations in terms of high-order cumulants of transport and their relationships.
Second, as we prove below, harmonic systems exactly satisfy the TUR (\ref{eq:TUR1}), 
making it clear that its validity extends beyond Markovian dynamics to cover more general cases.
%
Third, our simulations below satisfy the QTUR, yet we observe that 
this bound is quite loose, thus suggesting the existence of a tighter bound.
Altogether, as we illustrate in Secs. \ref{Sec-har}-\ref{Sec-ele}, invalidating the 
bound (\ref{eq:TUR0}) allows us to 
recognize systems of different statistics, as well as identify the onset of 
interactions and non-Markovianity in the dynamics, making it profoundly useful.

\section{Coupled harmonic oscillators: Proof for the Validity of the TUR}
\label{Sec-har}

We consider a noninteracting thermal transport model consisting 
of coupled harmonic oscillators.  
We assume that the first and last oscillators are coupled to independent heat baths, $L$ and $R$,
which are maintained at different temperatures, $T_L$ and $T_R$, respectively. 
The baths include a collection of harmonic oscillators, which  are bilinearly coupled to the system.
The model is fully harmonic and the total Hamiltonian for this setup can be written as
\begin{eqnarray}
\hat H&=&\sum_{\nu=L,R} \left(\!\frac{1}{2} \hat p_{\nu}^T \,\hat p_{\nu}\! + 
\!\frac{1}{2} \hat u_{\nu}^{T} K^{\nu} \hat u_{\nu} \right)
+ \!\frac{1}{2} \hat p_{C}^T \,\hat p_{C}\! + \!\frac{1}{2} \hat u_{C}^{T} K^{C} \hat u_{C} 
\nonumber\\
&+& \hat u_{L}^{T} V^{L C} \hat u_C + \hat u_{R}^{T} V^{R C} \hat u_C.
\end{eqnarray}
Here $\hat u_{\nu}$ and $\hat p_{\nu}$ are column vectors
of mass weighted coordinate operators and momenta
for regions $\nu=L,R$.  $K^{\nu}$ is the corresponding force constant matrix. 
Similar definitions hold for the central (C) region. 
$V^{LC}$ and $V^{RC}$ are the force constants matrices between the central system and the left and right regions;
recall that the left bath is coupled to a single (`first') oscillator in the system, and similarly the right bath is connected
to a specific (`last') oscillator.  
$T$ stands for the transpose operation. 

The integrated thermal energy current is 
defined as the net change of thermal energy in the reservoir, 
say the left one, ${Q}\equiv\hat H_L(0)- \hat H_L^H(t)$.
An exact expression for the steady state                 
CGF, ${\chi}_{\rm HO}(\alpha)$ for the integrated thermal energy current
can be obtained by following the two-time measurement protocol 
and employing the Keldysh non-equilibrium Green's function approach 
\cite{bijay-PRE-fcs, dharfcs07,saito-fluc}. 
The CGF is given by
\bea
&&\chi_{ HO}(\alpha)  = \!\!-\int_{-\infty}^{\infty} \frac{d\omega}{4 \pi} \ln \Big\{ 1 - {\cal T}_{HO}(\omega) \big[ n_L (\omega) \bar{n}_R(\omega) 
\times 
\nonumber \\&& (e^{i \alpha \hbar \omega}\!-\!1) + n_R(\omega) \bar{n}_L(\omega) (e^{-i \alpha \hbar \omega}\!-\!1) \big]\Big\},
\label{eq:CGFHO}
\eea
with the  transmission function ${\cal T_{HO}}(\omega)$, which is expressed
in terms of the Green's function of the central region and the self-energies of the baths 
\cite{bijay-PRE-fcs}. We highlight that the transmission function is derived for a harmonic system,
but the oscillators within could be coupled to each other with any geometry and force constants.
The baths are maintained at thermal equilibrium with the 
Bose-Einstein distribution function $n_{\nu}(\omega)= \left[\exp(\beta_{\nu} \hbar \omega) -1 \right]^{-1}$;
 $\bar{n}_{\nu}(\omega) \equiv 1 + n_{\nu}(\omega)$. 
Note that the transmission function does not depend on temperature
due to the noninteracting nature of the model.  
As a consequence, the above CGF satisfies the following symmetry 
\be
\chi_{HO}(\alpha;T_L, T_R) =\chi_{HO}(-\alpha; T_R, T_L),
\ee
which ensures that the energy current and the associated noise are odd and even functions 
of the affinity $\Delta \beta$, respectively. 
Therefore, for a fully harmonic model 
only even powers in $\Delta \beta$ survive in the TUR expression, Equation (\ref{eq:central-eq}). 

The analytical forms for the first and second cumulants 
can be readily obtained from Eq. (\ref{eq:CGFHO}),  and are given by
\begin{eqnarray}
\langle j \rangle &=& \int_{0}^{\infty} \frac{d\omega}{2 \pi} \hbar \omega {\cal T}_{HO}(\omega)  \big[n_L (\omega) - n_R(\omega)\big],\\
\langle j^2 \rangle_c &=& \int_{0}^{\infty} \frac{d\omega}{2 \pi} (\hbar \omega)^2 \Big\{ {\cal T}_{HO}(\omega) \big[n_L (\omega)  \bar{n}_R(\omega) \nonumber \\
&+& n_R(\omega)\bar{n}_L(\omega)\big] + {\cal T}_{HO}^2(\omega) \left[n_L(\omega)\!-\!n_R(\omega)\right]^2. \nonumber \\
\label{eq:jj2HO}
\end{eqnarray}
%
%
Expanding the current and the variance in powers of $\Delta \beta$, we construct the coefficient
$R_1= 3S_2-2G_3$, which is always positive,
\bea
R_1\!&=&\!\int_{0}^{\infty} \frac{d\omega}{2 \pi} (\hbar \omega)^4 {\cal T}_{HO}(\omega) n(\omega) \bar{n}(\omega) \Big[ 1 +
 \nonumber \\
&& 6 {\cal T}(\omega) n(\omega) \bar{n}(\omega)\Big] > 0.
\eea
Here $n(\omega)$ is the Bose-Einstein distribution function evaluated at the average temperature $T=(T_L+T_R)/2$.
Since the transmission function and the thermal distribution functions are positive,
the coefficient $R_1$ is always positive. 
Therefore, up to $\mathcal{O}(\Delta \beta)^2$ the TUR is satisfied for quantum harmonic networks
consisting of an arbitrary number of oscillators with general connectivity and 
arbitrary coupling strengths to the baths. 

Moreover, we now prove that for harmonic systems
the TUR is valid arbitrarily far from equilibrium.  
From Eq. (\ref{eq:jj2HO}), we note that the second cumulant obeys the inequality,
\begin{eqnarray}
\langle j^2 \rangle_c && \geq \int_{0}^{\infty} \frac{d\omega}{2 \pi} (\hbar \omega)^2 {\cal T}_{HO}(\omega) \big[n_L (\omega) \bar{n}_R(\omega) + n_R(\omega)\bar{n}_L(\omega)\big]. \nonumber \\
\end{eqnarray}
An equality is satisfied at thermal equilibrium, $\beta_L=\beta_R$. 
Interestingly, one can show the following inequality for $\forall \,\omega >0$,
\begin{equation}
\Big[n_L (\omega) \bar{n}_R(\omega) + n_R(\omega) \bar{n}_L(\omega)\Big] \geq \frac{2}{\Delta \beta \, \hbar \omega}\, \big[n_L(\omega) - n_R(\omega)\big].
\label{ineq}
\end{equation}
Using this inequality in the noise expression immediately implies that 
\begin{eqnarray}
\langle  j^2  \rangle_c &&\geq \frac{2}{\Delta \beta} \int_{0}^{\infty} \frac{d\omega}{2 \pi} \, 
\hbar \omega \, {\cal T}_{HO}(\omega) \, \left[n_L(\omega) - n_R(\omega)\right] \nonumber \\
&& = \frac{2}{\Delta \beta} \langle j \rangle, 
\end{eqnarray}
which is the TUR, Eq. (\ref{eq:TUR1}).  
This derivation is entirely independent of the details of the transmission 
function. The proof only emerges from the formal structure of the CGF in Eq. (\ref{eq:CGFHO}).
We conclude that for harmonic junctions the TUR is satisfied in the quantum and classical (high temperature) 
limits irrespective of the underlying dynamics, which could be Markovian or non-Markovian.
While this proof holds for classical and quantum systems alike,
in Appendix A we separately study classical harmonic systems by directly studying the classical CGF.

\section{Non-equilibrium Spin-Boson model: TUR violation for structured baths} 
\label{Sec-SB}

In the previous section, we proved the TUR, Eq. (\ref{eq:TUR1}) in harmonic systems.
It is intriguing to depart from the harmonic limit, which describes heat transport
in a system of independent normal modes, and examine the role of
anharmonicity (many-body interaction) for invalidating the TUR.
The nonequilibrium spin-boson model (NESB)
serves as a toy model for exploring the role of anharmonicity in quantum thermal transport.
Its transport characteristics have been extensively examined with different theoretical techniques;
partial list includes  \cite{SBDS,SB-thoss,SB-nic,SB-Cao,SB-NJP,SB-JJ,SB-Kil,SB-Kelly}.

The NESB model comprises a two-level system (spin) interacting with two bosonic environments, 
$\nu=L,R$, which are maintained at different temperatures. 
The Hamiltonian for the NESB model reads 
\bea
\hat H = \frac{\hbar \omega_0}{2} \hat \sigma_z +  \frac{\hbar \Delta}{2} \hat \sigma_x +
\sum_{\nu,j}\omega_{\nu,j} \hat b_{\nu,j}^{\dagger}\hat b_{\nu,j} \nonumber \\+
\hat \sigma_z \sum_{\nu,j}\hbar \gamma_{\nu,j}(\hat b_{\nu,j}^\dagger + \hat b_{\nu,j}).
\label{eq:HSB}
\eea
Here, $\hat b_{\nu,j}^{\dagger}$ ($\hat b_{\nu,j}$) are bosonic creation (annihilation) operators
of the $j$th mode in the $\nu$ thermal bath. 
$\omega_0$ is the spin splitting, $\Delta$ the tunneling element,
$\hbar \gamma_{\nu,j}$ the coupling energy of the spin polarization to the displacements of baths' oscillators. 
Given the quantum nature of the spin system and the varied applicability of the model, 
understanding the validity of the thermodynamic uncertainty relation in the NESB is of a significant  interest. 


To examine the possible violation of the TUR in the 
NESB model, we need to study it beyond the weak-coupling Markovian dynamics \cite{hava}.
%
The nonequilibrium Green's function method in combination 
with the Majorana fermion representation for the spin \cite{SB-bijay}
extends beyond the Markovian weak coupling limit, thus it may allow to observe TUR violation.
Using the Majorana Green's function  approach, we derived  in Ref. \cite{SB-bijay} 
the CGF for the NESB model, where for simplicity,  we focus on the unbiased case, $\omega_0=0$,
\bea
{\chi}_{SB}(\alpha) \!&\!=\!\!&\int_{-\infty}^{\infty}\! \frac{d\omega}{4 \pi}\, \frac{\Delta^2}{\omega^2}\, \! \ln  
\Big\{ 1 + {\cal T}_{SB}(\omega; T_L, T_R) \Big[ n_L(\omega)\bar{n}_R(\omega)  
\nonumber \\
&&( e^{i \alpha \hbar \omega}\!-\!1) + n_R(\omega)\bar{n}_L(\omega) (e^{-i \alpha \hbar \omega}\!-\!1)\Big] \Big\}.
\label{eq:CGF-SB}
\eea
The transmission function of the NESB model is given by \cite{SB-bijay}
\begin{small}
\be
{\cal T}_{SB}(\omega; T_L,T_R) = 
\frac{4 \,\Gamma_L(\omega) \,\Gamma_R(\omega) \,\omega^2}{ \Big(\omega^2 \!-\!\Delta^2\Big)^2 + \omega^2 \,  
\Big[\sum_{\nu}\Gamma_{\nu}(\omega) (1\!+\! 2 n_{\nu}(\omega))\Big]^2}.
\label{eq:SB-trans}
\ee
\end{small}
Note the crucial sign difference between the CGF for the spin-boson model in comparison 
to the harmonic case, Eq. (\ref{eq:CGFHO}).
Here, $\Gamma_{\nu}(\omega)=2\pi \sum_{j} \gamma_{\nu,j}^2 \delta (\omega-\omega_{\nu,j})$ is the 
spectral function of the $\nu$ bath. 
As a reflection of the model's anharmonicity, the transmission function depends on temperature, 
which results in
\be
\chi_{SB}(\alpha;T_L, T_R) \neq \chi_{SB}(-\alpha; T_R, T_L).
\ee
Therefore, odd powers of $(\Delta \beta)$ contribute to the TUR expression (\ref{eq:central-eq}).
From Equation (\ref{eq:CGF-SB}), we obtain an analytical expression for the coefficient $R_1$, 
\bea
R_1= \int_{-\infty}^{\infty} \frac{d\omega}{4 \pi} (\hbar^3\Delta^2 \omega) {\cal T}_{SB}(\omega;T) \left(-\frac{\partial n(\omega)}{\partial \beta}\right)
\nonumber \\ 
\times \Big[ 1 - 6 {\cal T}_{SB}(\omega;T) n(\omega) (1+n(\omega))\Big]. 
\label{eq:R1SB}
\eea
Here, ${\cal T}_{SB}(\omega;T)$ is evaluated at the average temperature $T=(T_L+T_R)/2$.
Once again note the sign difference in $R_1$ in comparison to the harmonic case, reflecting the hybrid nature (spin-boson) of the system. The sign of $R_1$ can therefore be negative, which could result in the violation of the TUR, Eq. (\ref{eq:TUR1}).

\begin{figure}
\begin{center}
\includegraphics[width=\columnwidth]{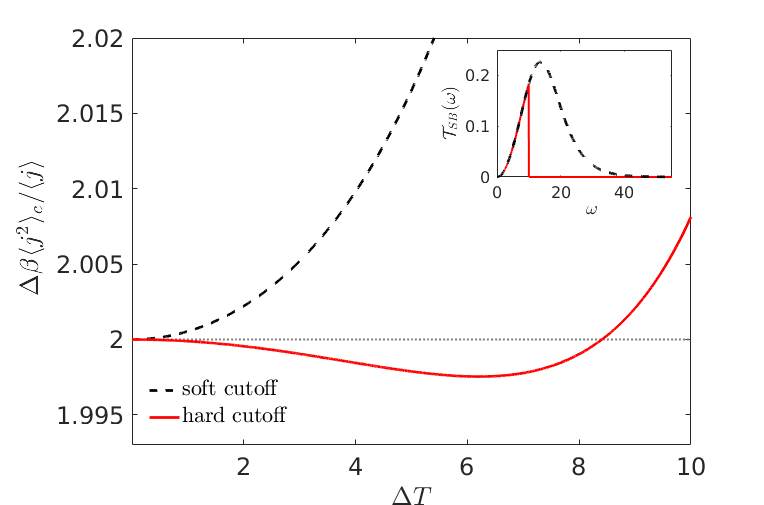} 
\caption{TUR in the NESB model based on Eqs. (\ref{eq:CGF-SB})-(\ref{eq:SB-trans}).
The TUR holds when the ohmic spectral density function has a soft cutoff (dashed), 
but is invalidated within a certain window when the cutoff is hard (full).
Other parameters are $T=10$, 
$\omega_c=10$, $\Delta=1$, $\kappa=0.5$.
The inset present the transmission function for soft (dashed) and hard (full) cutoff at $\Delta T =5$,
where in the latter case it immediately falls to zero at $\omega_c=10$.
}
\label{fig:TUR-SB}
\end{center} 
\end{figure}

In Fig. \ref{fig:TUR-SB}, we present numerical simulations of the combination
$\Delta \beta \frac{\langle j^2 \rangle_c}{  \langle j\rangle}$
in the NESB model using the Majorana NEGF method.
We find that the TUR can be violated by structuring the spectral density function 
of the thermal reservoirs. Specifically, we used an Ohmic function with either a soft cutoff,
$\Gamma_{\nu}(\omega)=\kappa \pi \omega e^{-|\omega|/\omega_c}$,
or a hard cutoff, $\Gamma_{\nu}(\omega)=\kappa \pi \omega e^{-|\omega|/\omega_c} \Theta (\omega_c-|\omega|)$;
the exponential function plays a minor role in the latter expression, but we keep it so as to maintain
a quantitative comparison between the two models.
In both cases, we used $\omega_c=10\Delta$ as the cutoff frequency.
This value seems to be high, thus at first sight it suggests the Markovian limit.
However, a closer look (inset) reveals that in fact 
the transmission function comprises significant weight beyond $\omega_c$ 
in a region where $n_{\nu}(\omega)$ is still substantial,
given the high temperature employed.

Figure \ref{fig:TUR-SB} shows that the TUR is obeyed in the case of
a soft-cutoff spectral function.
In contrast, we observe a weak but noticeable violation of the TUR once we structure 
the baths and filter high frequencies using a hard cutoff.
This effect can be attributed to the fact that transport is non-Markovian in this limit since the baths comprise modes
only at or below the averaged temperature.
Nevertheless, a fundamental understanding of the TUR violation in the NESB model in the language  of the underlying
transport mechanism is still missing.

It is instructive to analyze the TUR in the NESB model in two special cases: 

\noindent{\it Weak coupling limit.}
When the system-bath coupling is weak, the transmission function (\ref{eq:SB-trans}), evaluated
at $T=T_L=T_R$,
can be approximated by the Dirac delta function,
\begin{equation}
{\cal T}_{SB}(\omega;T)=
\frac{\Gamma(\Delta) \, \pi}{[1+2n(\Delta)]}[\delta(\omega+\Delta)+\delta(\omega-\Delta)].
\end{equation}
The spectral function is evaluated at the tunneling energy $\Delta$, and 
for simplicity we assume that it is identical at the two contacts, $\Gamma(\Delta)=\Gamma_{\nu}(\Delta)$.
Similarly, the Bose-Einstein distribution is evaluated at the spin splitting $\Delta$,
$n(\Delta)=[e^{\Delta/k_BT}-1]^{-1}$.
The weak coupling limit is also known as the sequential (or resonant) tunneling limit.
We further calculate,  in the weak coupling limit,
the following function,
\begin{equation}
 {\cal T}_{SB}^2(\omega;T)=\frac{\Gamma(\Delta) \, \pi}{2[1+2n(\Delta)]^3}[\delta(\omega+\Delta)+\delta(\omega-\Delta)].
\end{equation}
Substituting these expressions into $R_1$,  Eq. (\ref{eq:R1SB}),  we get
\begin{equation}
R_1=\frac{\hbar^4\Delta^4 \, \Gamma(\Delta) \, n(\Delta)[1+n(\Delta)]}{2[1+2n(\Delta)]^3}[n(\Delta)^2+n(\Delta)+1],
\label{eq:R1SBweak}
\end{equation}
which is always positive. 
The same conclusion is reached by following the CGF obtained using the Markovian Redfield 
quantum master equation approach \cite{SB-nic,hava}, see Appendix B.
Overall, this proves that in the weak coupling limit, 
there is no violation to the TUR in the quadratic order, $(\Delta \beta)^2$.
It remains a challenge to prove (or disprove) the validity of the TUR arbitrarily 
far from equilibrium in the weak coupling limit.

\noindent{\it Co-tunneling limit.} 
At low temperatures, $\Gamma<T_{\nu}<\Delta$, sequential tunneling is exponentially suppressed
since incoming phonons fall short of the spin splitting $\Delta$. The residual off-resonant
transmission probability is therefore small,
${\mathcal T}_{SB}(\omega)\ll 1$. 
From the CGF (\ref{eq:CGF-SB}) we find the current and its noise
in this co-tunneling limit \cite{SB-bijay},
\bea
\langle j \rangle &=& \frac{2}{\pi} \int_{0}^{\omega_h} d\omega \hbar \omega {\cal T}_{\rm co}(\omega) 
\big[n_L(\omega) - n_R(\omega)\big],
\nonumber \\
\langle j^2\rangle_c &=& \frac{2}{\pi} \int_{0}^{\omega_h} d\omega (\hbar \omega)^2 {\cal T}_{\rm co}(\omega) 
\big[n_L(\omega) \bar{n}_R(\omega) + n_R(\omega)\bar{n}_L(\omega)\big] \nonumber \\
\eea
Here,  the upper limit of the integral $\omega_h$ is determined by the smallest of two energy scale:
temperature or the cutoff frequency of the baths.
$\mathcal{T}_{\rm co}(\omega)$ stands for the transmission function in the co-tunneling limit
\cite{SB-bijay},
\bea
\mathcal{T}_{\rm co}(\omega) = \frac{\Gamma_L(\omega)\Gamma_R(\omega)}{\Delta^2},
\eea
which is independent of temperature. 
Following the inequality (\ref{ineq}), 
it is straightforward to show that the TUR is satisfied in the co-tunneling limit.
In fact, the NESB model behaves similarly to the harmonic junction in the co-tunneling limit since $\Delta>T_{\nu}$.
Therefore, it is not surprising that the TUR is satisfied in this case.

\section{Thermal transport of noninteracting electrons: TUR violation in the resonant transport regime}
\label{Sec-ele}

After analyzing the validity and breakdown of the TUR for bosonic heat transport and for the NESB model, 
in this Section we study electronic heat transport.
For simplicity, we focus on a one-dimensional noninteracting tight-binding chain model for fermions
with the Hamiltonian 
\bea
\hat H= \sum_{\nu=L,C,R} 
\hat c_{\nu}^{\dagger} h^{\nu} \hat c_{\nu} + 
\sum_{\nu=L,R}  \Big(\hat c_{\nu}^{\dagger} V_{e}^{\nu C} \hat c_C + {\rm h.c.}\Big).
\eea
Here, $\hat c^{\dagger}_{\nu} (\hat c_{\nu})$ is the row (column) vector consisting of 
electronic creation (annihilation) operators in the $\nu$ region, with
$h^{\nu}$ the single-particle Hamiltonian matrix in that domain.
$V_e^{\nu C}$ is the coupling matrix between the metals and the central system. 
The baths ($\nu=L,R$) and the central system are initially decoupled and are 
prepared at their respective grand canonical equilibrium state with temperature $T_{\nu}$ 
and chemical potential $\mu_{\nu}$. 

The steady state expression for the joint CGF corresponding to integrated particle and energy current 
can be obtained exactly for this model. 
It was first derived by Levitov and Lesovik \cite{Levitov1, Levitov2} following a wave scattering 
approach. It was later derived by employing different rigorous approaches 
\cite{Schonhammer1, Schonhammer2, Klich, Esposito-review, Bernard,bijay-PRE-fcs}. The joint CGF is given as 
\bea
&&\chi_{el}(\alpha_p, \alpha_e)\!= 
\nonumber\\
&&\!\int_{-\infty}^{\infty}\! \!\frac{d\epsilon}{2 \pi \hbar} 
\ln \Big\{ 1 + {\cal T}_{el}(\epsilon) \big[ f_L (\epsilon) \bar{f}_R(\epsilon) (e^{i (\alpha_p + \epsilon \alpha_e)}\!\!-\!\!1) \nonumber \\
&&+ f_R(\epsilon) \bar{f}_L(\epsilon) (e^{-i (\alpha_p + \epsilon \alpha_e)}\!\!-\!\!1) \big]\Big\}.
\label{eq:CGF-el}
\eea
Here, $\alpha_p (\alpha_e)$ is the counting parameter keeping track of the net particle 
(energy) exchange between the system and the left bath in steady state. 
$f_{\nu}(\epsilon)= 1/\big[e^{\beta_{\nu}(\epsilon-\mu_{\nu})}+1\big]$ 
is the Fermi-Dirac distribution function for the leads,  $\nu=L,R$
$\bar{f}_{\nu}(\epsilon) = 1 - f_{\nu}(\epsilon)$, 
$\mu_{\nu}$ is the corresponding chemical potential and ${\cal T}_{el}(\epsilon)$ is the transmission function,  containing the structural information of the electrodes and the system. 
The joint CGF satisfies the steady state fluctuation symmetry, 
$\chi_{el}(\alpha_p, \alpha_e) = \chi_{el}(-\alpha_p + i (\beta_L \mu_L -\beta_R \mu_R), -\alpha_e+i \Delta \beta)$.

Following Eq. (\ref{eq:CGF-el}), the CGF for heat exchange can be obtained by simply replacing 
$\alpha_p= -\mu_L \alpha$ and $\alpha_e = \alpha$, 
where $\alpha$ keeps track of net amount heat transfer at the left contact. 
We then receive
\bea
\chi_{el}(\alpha) &=& \int_{-\infty}^{\infty} \frac{d\epsilon}{2 \pi \hbar} 
\ln \Big\{ 1 + {\cal T}_{el}(\epsilon) \big[ f_L (\epsilon) \bar{f}_R(\epsilon) (e^{i \alpha (\epsilon -\mu_L)}\!-\!1) \nonumber \\
&+& f_R(\epsilon) \bar{f}_L(\epsilon) (e^{-i \alpha (\epsilon-\mu_L)}\!-\!1) \big]\Big\}.
\label{eq:CGF-el-heat}
\eea
Since our focus here is on the heat current, we consider identical chemical potentials 
($\mu =\mu_R =\mu_L$) but different temperatures ($T_L \neq T_R)$ for the leads. 
In this case, the CGF follows the symmetry 
$\chi_{el}(\alpha) = \chi_{el}(-\alpha + i \Delta \beta)$. Note that, similarly 
to the noninteracting bosonic case, the CGF here also satisfies
\be
\chi_{el}(\alpha;\mu, T_L, T_R) =\chi_{el}(-\alpha;\mu, T_R, T_L),
\ee 
which implies that the TUR expression in Eq.~(\ref{eq:central-eq}) contains only 
even powers in  $\Delta \beta$. 
Using Eq. (\ref{eq:CGF-el-heat}), the heat current and the associated noise are given by
\bea
\langle j \rangle &=& \int_{-\infty}^{\infty} \frac{d\epsilon}{2 \pi \hbar} (\epsilon-\mu) {\cal T}_{el}(\epsilon)  \big[f_L (\epsilon) - f_R(\epsilon)\big], 
\label{eq:je}
\\
\langle j^2 \rangle_c &=& \int_{-\infty}^{\infty} \frac{d\epsilon}{2 \pi \hbar}(\epsilon-\mu)^2 \Big\{ {\cal T}_{el}(\epsilon) \big[f_L(\epsilon)  \bar{f}_R(\epsilon) \nonumber \\
&+& f_R(\epsilon)\bar{f}_L(\epsilon)\big] - {\cal T}^2_{el}(\epsilon) (f_L(\epsilon)-f_R(\epsilon))^2\Big\}.
\label{eq:je2}
\eea
As before, we compute $R_1$, which is given by,
\bea
R_1 &=&\int_{-\infty}^{\infty} \frac{d\epsilon}{2 \pi \hbar} (\epsilon-\mu)^4 \, {\cal T}_{el}(\epsilon)\,  f(\epsilon)
 [1-f(\epsilon)] \times 
\nonumber \\
&&\Big[ 1 - 6 {\cal T}_{el}(\epsilon) f(\epsilon) [1-f(\epsilon)]\Big].
\label{eq:R1_expression}
\eea
The Fermi function $f(\epsilon)$ is evaluated at the averaged temperature.
It is important to note that the above obtained expressions are valid for arbitrary temperatures, 
bias voltage, system-bath coupling and the details of the chain, encapsulated within
the transmission function.
In what follows, we once again consider different limiting cases to analyze the TUR. 


\begin{figure}[htbp]
\includegraphics[width=\columnwidth]{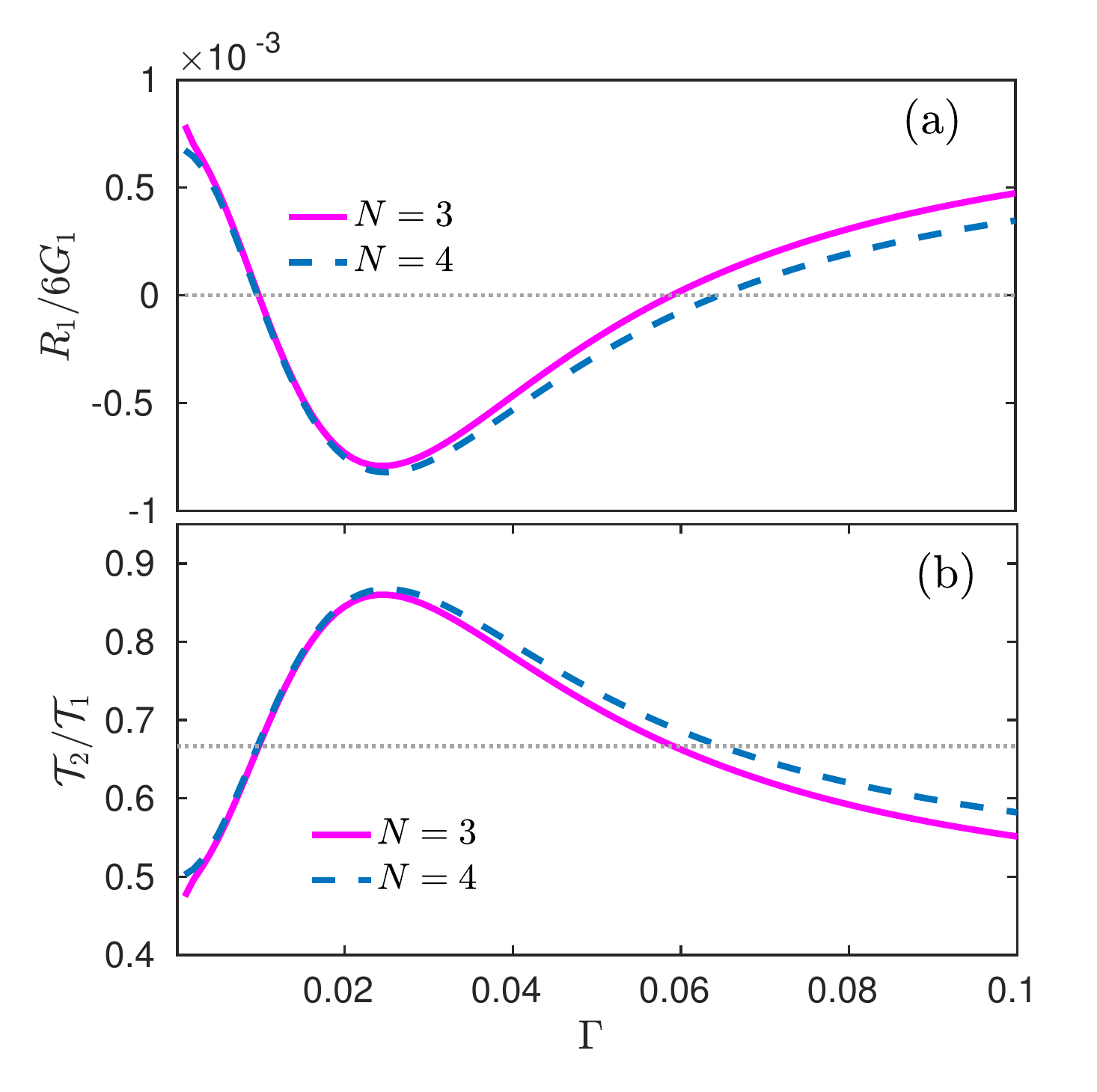} 
\caption{TUR violation in electronic thermal transport junctions.
(a) Simulation of $R_1$ using Eq. (\ref{eq:R1_expression}), demonstrating TUR violation
within a certain range for $\Gamma$ for 
$N=3$ (full) and $N=4$ (dashed).
(b) This violation is in accordance
with the observation of  ${\cal T}_2/{\cal T}_1 > 2/3$,
where the dotted line marks the value $2/3$.
Parameters are $\beta=1/4$, $\Omega=1/80$, $\mu=0$.
$\epsilon_d=1/8$.}
\label{FigE1}
\end{figure}

\begin{figure}[htbp]
\includegraphics[width=\columnwidth]{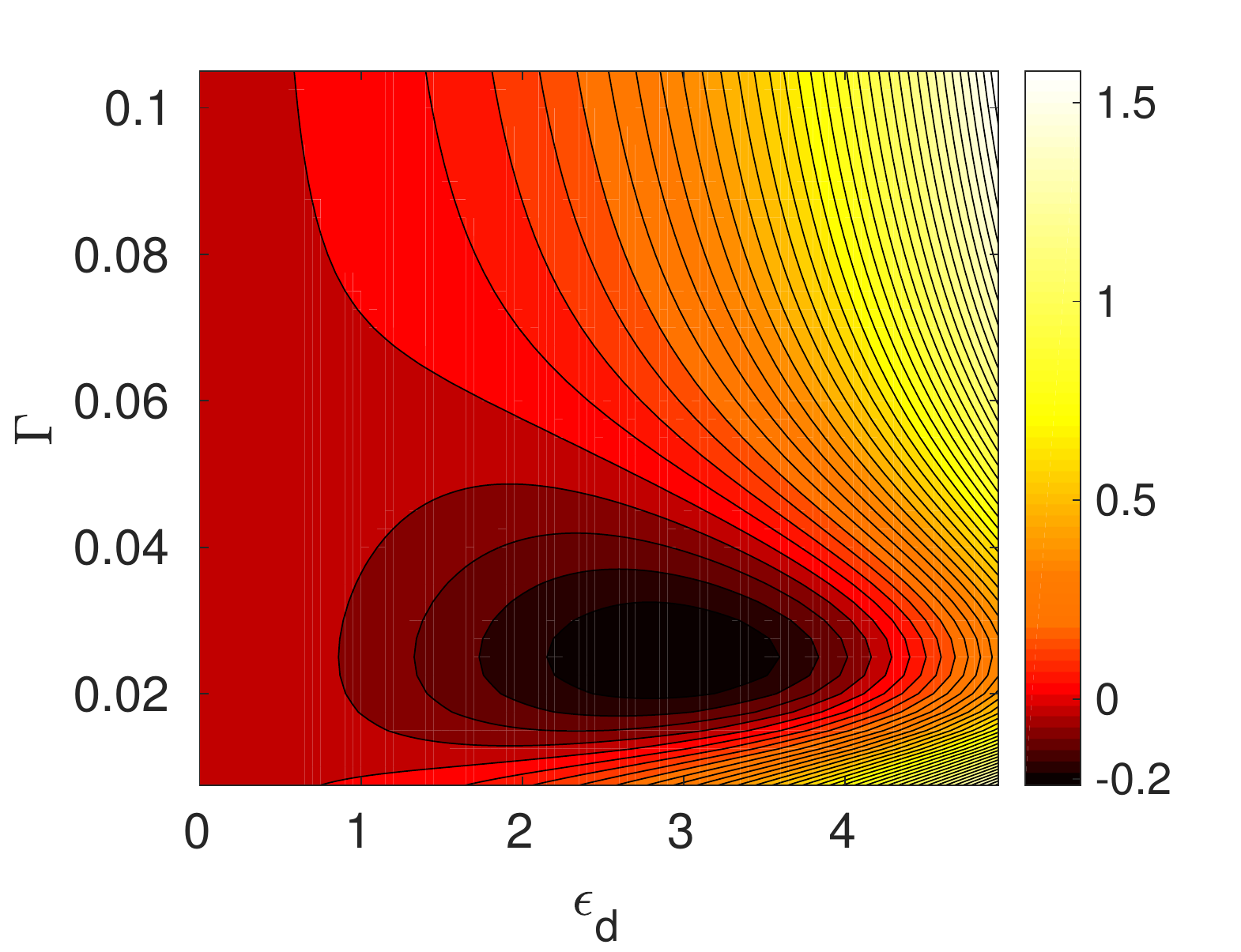}
\caption{Contour plot of the ratio $R_1/6G_1$ in electronic thermal transport junctions
as a function of $\epsilon_d$ and $\Gamma$, beyond the resonant tunneling limit.
The TUR is violated (to the order $(\Delta \beta)^2$) when $R_1<0$.
Other parameters are $\beta=1/4$, $\Omega=1/80$, $\mu=0$, $N=3$.
}
\label{FigE2}
\end{figure}

\noindent {\it Low-transmission.} 
In the limit when ${\cal T}_{el}(\epsilon)\ll 1$ for all values of $\epsilon$, one can discard 
the term ${\cal T}_{el}^2(\epsilon)$ in Eq.~(\ref{eq:R1_expression})
relative to ${\cal T}_{el}(\epsilon)$. This immediately implies 
that the TUR is satisfied.

Indeed, in the context of charge transport, low transmission probability 
is associated with a Poisson process, which further implies 
uncorrelated electron transport through the junction. 
In this limit, one simplifies Eq.~(\ref{eq:CGF-el-heat}) by approximating $\ln (1 + x) \approx x$ 
and writes the CGF as 
\bea
\chi_{el}(\alpha) &=& \int_{-\infty}^{\infty} 
\frac{d\epsilon}{2 \pi \hbar} {\cal T}_{el}(\epsilon) \big\{ f_L (\epsilon) \bar{f}_R(\epsilon) (e^{i \alpha (\epsilon -\mu_L)}\!-\!1) \nonumber \\
&+& f_R(\epsilon) \bar{f}_L(\epsilon) (e^{-i \alpha (\epsilon-\mu_L)}\!-\!1) \big]\Big\}.
\eea
%
In Appendix C we further prove that the TUR is valid  arbitrarily far from equilibrium in this low transmission transport regime.

\noindent{\it Constant transmission.} 
If the transmission is a constant i.e., ${\cal T}_{el}(\epsilon)= \tau$, 
the integration in Eq.~(\ref{eq:R1_expression}) can be performed analytically and the expression for TUR up to $\mathcal{O}{(\Delta \beta)^2}$ simplifies to
\bea
\Delta \beta \frac{\langle j^2\rangle_c}{\langle j \rangle}= 2 +  \frac{(\Delta \beta)^2}{\beta^2} \Big[ \frac{7}{5} \pi^2 (1-\tau) + 12 \tau\Big]\geq0,
\eea
which once again validates the TUR. For perfect transmission $\tau=1$, the above expression simplifies to $\Delta \beta \frac{\langle j^2\rangle_c}{\langle j \rangle}= 2 + 12 \frac{(\Delta \beta)^2}{\beta^2} \geq 0$. 

\noindent{\it Weak coupling- Resonant tunneling limit.}
Another experimentally realizable  case
is the weak system-bath coupling limit, also known as the sequential tunneling or the resonant tunneling regime. 
In this case, the transmission function is sharply peaked about resonance frequencies 
(corresponding to molecular/quantum dot electronic levels),
while the Fermi functions are relatively broad (constant) 
in the range where the transmission function
is non-zero i.e., $k_B T_{\nu} > \Gamma, \epsilon_d$, where $\Gamma$ is the hybridization energy and $\epsilon_d$ the
characteristic energy level of the quantum dots.
As a precaution we note that while for charge transport we assess the width of the transmission function
itself---relative to the Fermi function---for thermal energy transport
we need instead to confirm that the combined function $(\epsilon-\mu)^4{\cal T}_{el}(\epsilon)$ is sufficiently narrow
relative to the alteration of the Fermi functions. 

Assuming for simplicity that the central region includes electronic resonances  
clustered around the energy $\epsilon_d$, Eq.~(\ref{eq:R1_expression})  simplifies to 
\bea
R_1 = f(\epsilon_d)[1-f(\epsilon_d)] {\cal T}_1 - 6 f(\epsilon_d)^2 [1-f(\epsilon_d)]^2  {\cal T}_2,
\label{eq:R1elweak}
\eea
where we define
\bea
{\cal T}_n = \int_{-\infty}^{\infty} \frac{d\epsilon}{2 \pi \hbar} (\epsilon-\mu)^4 {\cal T}_{el}^n(\epsilon).
\label{integrals}
\eea
These integrals converge if the transmission ${\cal T}_{el}(\epsilon)$ 
decays faster that $1/\epsilon^4$.
Based on Eq. (\ref{eq:R1elweak}) and the inequality $ 0 \leq f(1-f) \leq 1/4$,
violation of the TUR  ($R_1<0$) occurs when
\bea
\frac{{\cal T}_2}{{\cal T}_1}> \frac{2}{3}.
\label{eq:T2T1}
\eea
This inequality can be satisfied 
by tuning the electronic parameters of the chain and its hybridization to the metals,
as we show below.



In Figs. \ref{FigE1}-\ref{FigE2} we present numerical results for the TUR in electronic heat transport following Eq.~(\ref{eq:R1_expression}). 
The model consists a junction with its central part including quantum dots in a serial configuration, 
and we set the spectral function for the baths to be constant and identical,
$\Gamma = \Gamma_L = \Gamma_R$.
The $N$ serial quantum dots are described by a tight-binding model
with onsite energy $\epsilon_d$ and hopping parameter $\Omega$.
The transmission function is calculated in a standard way 
from the Green's function of the system and 
self energy of the baths \cite{Nitzan,diventra}.

In Fig. \ref{FigE1}(a),
we display the ratio $R_1/6G_1$ as a function of $\Gamma$ 
and demonstrate the violation ($R_1/6G_1<0$) of the TUR within a certain range of this coupling. We present  results for $N=3$ and $N=4$. Interestingly, the violation weakly depends on the number of quantum dots in the setup,
and in fact for the present parameters, results saturate beyond $N=4$.
We further show in Fig.~\ref{FigE1}(b) that the condition of Eq. (\ref{eq:T2T1}),  
which was derived under the assumption of resonant tunneling,
very well captures the violation region. 
Note that for $N=3$ and $N=4$ the transmission function ${\cal T}_{el}(\epsilon)$ for large $\epsilon$ decays as $1/\epsilon^6$ and $1/\epsilon^8$ respectively and therefore the integrals in Eq.~(\ref{integrals}) converge.

Figure \ref{FigE2} displays a map of $R_1/6G_1$, while extending beyond the resonant tunneling limit,
with $\epsilon_d$ exceeding $T$.
Negative values correspond to the breakdown of the TUR, and we
identify a significant basin of TUR violation  when $\Gamma\ll\epsilon_d<T$.

\section{Summary}
\label{Sec-summ}

We used the steady state fluctuation symmetry to explore
the validity of the TUR in thermal transport problems.
From the SSFS, we wrote down relationships between transport coefficients
and organized an expression for the TUR, which was perturbative in
the affinity $\Delta \beta$, and given in terms of nonlinear transport coefficients. 
The first central result of this work is that 
negative skewness (to the lowest order in the perturbative expansion around equilibrium) 
reveals TUR violations.
This result is consistent with our previous work on charge transport \cite{BijayTUR}.
%
Our expansion, building the TUR, is universally valid for
quantum and classical problems, as well as for arbitrary interactions 
in the conducting system.

The second important result of our work is the proof that
the TUR is satisfied in harmonic junctions, classical and quantum,
irrespective of the underlying stochastic dynamics.
Furthermore, to understand the impact of interaction on the violation of the TUR
we studied the nonequilibrium spin-boson model. For this model we showed that
the TUR can be violated if the phonon baths are structured by employing a hard frequency
cutoff for their spectral density function, thus eliminating
a significant portion of the transmission function.

Finally, we studied the TUR in fermionic chains. 
Here, we focused on the resonant transport limit where an analytical condition
for TUR violation was derived, demonstrated to be in a quantitative agreement
with simulations. The TUR was satisfied when electron transmission probability was small (Poissonian statistics).
Violations were identified within a certain range of the system-bath hybridization energy.


The thermodynamic uncertainty relation attracts significant interest 
given its impact on the performance of thermal machines.
Unlike our previous study, which was focused on quantum charge transport and assumed noninteracting electrons,
the present work deals with heat exchange in classical and quantum systems, and it exposes that the validity of the TUR depends 
on the statistics of the participating particles as well as on their interaction.
For bosonic systems, interactions are necessary to invalidate the TUR, 
while fermionic chains can show violations even in the case of noninteracting electrons.
Future work will focus on the behavior of the TUR in 
classical anharmonic chains to understand the impact of quantum effects and many-body
interaction on fluctuation-entropy production trade-off relations.

\begin{acknowledgments}
The work of HMF was supported by the NSERC PGS-D program.
DS acknowledges the NSERC Discovery Grant and the Canada Chair Program. 
BKA would like to thank G. Schaller, G. Landi and G. Guarnieri for useful discussions. 
BKA gratefully acknowledges the start-up funding from IISER Pune, the Max Planck-India mobility grant and 
the hospitality of the Department of Chemistry at the University of Toronto.
\end{acknowledgments}

\renewcommand{\theequation}{A\arabic{equation}}
\setcounter{equation}{0}  

\section*{Appendix A: Harmonic junctions in the classical limit}

In this Appendix, we prove the validity of TUR directly in the classical limit for harmonic systems
starting from the classical generating function for heat exchange obtained by Kundu et al. \cite{classical-fluc}. 
One can as well reach the classical result from the exact quantum CGF in 
Eq.~(\ref{eq:CGFHO}) by taking the high temperature limit, 
$ \beta_L \hbar \omega \ll 1, \beta_R \hbar \omega \ll 1 $. 
The CGF in the classical limit, $\chi^{\rm cl}_{\rm HO}(\alpha)$, is given by, 
\bea
\chi^{\rm cl}_{HO}(\alpha)
= - \int_{-\infty}^{\infty} \frac{d\omega}{4 \pi} \ln \Big[ 1 \!-\! {\cal T}_{HO}(\omega) \frac{i \alpha}{\beta_L \beta_R} \big( i \alpha \!+\! (\beta_R\!-\!\beta_L)\big)\Big]\nonumber \\
\eea
Of course, this classical version also satisfies the steady state fluctuation relation 
i.e., $\chi^{ \rm cl}_{HO}(\alpha) = \chi^{\rm cl}_{\rm HO}(-\alpha + i \Delta \beta)$.
Using this CGF, one can immediately derive the current and its noise,
\bea
\langle j \rangle &\equiv& 
\frac{\partial \chi_{HO}^{\rm cl}(\alpha)}{\partial (i \alpha)}\Big{|}_{\alpha=0} =\,k_B(T_L\!-\!T_R) \, {\cal T}_1,
\nonumber \\
\langle j^2 \rangle_c &\equiv& \frac{\partial^2 \chi_{HO}^{\rm cl}(\alpha)}{\partial (i \alpha)^2}\Big{|}_{\alpha=0} = 2k_B^2 T_L T_R {\cal T}_1 + k_B^2(T_L \!-\!T_R)^2 {\cal T}_2,
\nonumber\\
\eea
where ${\cal T}_n = \int_{-\infty}^{\infty} \frac{d\omega}{4\pi} \, {\cal T}_{HO}^n (\omega)$. 
Since the last term in the noise expression is positive, one can then write
\bea
\langle j^2 \rangle_c \geq 2 \, T_L T_R {\cal T}_1 
=2 \frac{\langle j \rangle}{{\Delta \beta}},
\eea
which recovers the TUR (\ref{eq:TUR1}).The equality is reached in the equilibrium limit.
The TUR is therefore satisfied for a coupled harmonic oscillator system in the classical limit.
The proof is general for arbitrary spectral function of the baths and internal parameters for the system.
Thus, the TUR is valid independent of the nature of the underlying stochastic dynamics of the oscillators.


\renewcommand{\theequation}{B\arabic{equation}}
\setcounter{equation}{0}  

\section*{Appendix B: The nonequilibrium spin-boson model in the sequential tunneling limit}
We obtain here the weak coupling expression for $R_1$, Eq. (\ref{eq:R1SBweak}), 
directly from the weak-coupling CGF. 
The CGF of the NESB model in the sequential tunneling limit (weak system-bath coupling) 
was derived in Ref. \cite{SB-nic} 
using the Redfield quantum master equation approach. It was also received
as a special case following the Majorana Green's function technique \cite{SB-bijay},
\bea
\chi^{{\rm redfield}}_{{SB}}(\alpha)=-\frac{1}{2}[C(\Delta)-\sqrt{C^2 (\Delta) + 4 A(\Delta,\alpha)}],
\nonumber\\
\eea
where
\begin{eqnarray}
C(\Delta) &=&\Gamma_L(\Delta)[1+2n_L(\Delta)]+\Gamma_R(\Delta)[1+2n_R(\Delta)] 
\nonumber \\
A(\Delta,\alpha) &=& \Gamma_L(\Delta) \Gamma_R(\Delta)\big[n_L(\Delta) \bar{n}_R(\Delta) (e^{i \alpha \hbar \Delta}-1) \nonumber \\
&& + n_R(\Delta) \bar{n}_L(\Delta) (e^{-i \alpha \hbar \Delta}-1) \big].
\end{eqnarray}
%
A quick way to get the expression for $R_1$ is to obtain the third cumulant of
the current $\langle j^3 \rangle_c =\frac{\partial^3 \chi_{{\rm SB}}^{{\rm redfield}}}{\partial (i\alpha)^3}\bigg|_{\alpha=0}$ and  
perform a linear response analysis. Following that, we receive
\begin{equation}
R_1=\frac{\hbar^4 \Delta^4 \, \Gamma(\Delta) \, n(\Delta)[1+n(\Delta)]}{2 [1+2n(\Delta)]^3}[n(\Delta)^2+n(\Delta)+1]
\end{equation}
which is always positive. Here $n(\Delta)=\big[\exp(\beta \hbar \Delta)-1\big]^{-1}$  is the Bose-Einstein distribution function at temperature $T = 1/k_B \beta$. 
This expression matches exactly with Eq.~(\ref{eq:R1SBweak}). 


\vspace{10mm}
\renewcommand{\theequation}{C\arabic{equation}}
\setcounter{equation}{0}  

\section*{Appendix C: Validity of TUR for electronic heat transport in the low-transmission limit}
We prove here that the TUR relation for non-interacting electronic heat transport is valid at arbitrary far from equilibrium in the low transmission limit. We start from Eqs. (\ref{eq:je})-(\ref{eq:je2}), shift the energy around $\mu$, 
and take the limit of $\mathcal T_{el}(\epsilon)\ll 1 $,
\bea
\langle j \rangle &=& \int_{0}^{\infty} \frac{d\epsilon}{2 \pi \hbar} \epsilon \Big[{\cal T}_{el}(\epsilon \!+\! \mu) \!+\! {\cal T}_{el}(\epsilon \!-\! \mu) \Big] \big[f_L (\epsilon) \!-\! f_R(\epsilon)\big],
\nonumber \\
\langle j^2 \rangle_c &=& \int_{0}^{\infty} \frac{d\epsilon}{2 \pi \hbar} \epsilon^2 \Big[{\cal T}_{el}(\epsilon + \mu) + {\cal T}_{el}(\epsilon - \mu) \Big] \big[f_L(\epsilon)  \bar{f}_R(\epsilon) 
\nonumber \\
&+& f_R(\epsilon)\bar{f}_L(\epsilon)\big].
\eea
We notice the following inequality involving the Fermi functions for $\epsilon \geq 0$.
\bea
\Big[f_L (\epsilon) \bar{f}_R(\epsilon) \!+\! f_R(\epsilon) \bar{f}_L(\epsilon)\Big] \!&=&\! 
\coth\left(\frac{\Delta \beta \epsilon}{2}\right)
\left[f_L(\epsilon)\!-\!f_R(\epsilon)\right], \nonumber \\
 \geq && \!\!\!\!\!\frac{2}{\Delta \beta \, \epsilon }\,\big[f_L(\epsilon)\!-\!f_R(\epsilon)\big].
\label{eq:ineq-fermi}
\eea
Introducing this inequality in the above noise expression immediately implies that $\Delta \beta \frac{\langle j^2\rangle_c}{\langle j \rangle} \geq 2$, completing our proof on the validity of the TUR in the limit of low transmission probability.

\end{document}